\documentclass[pra,prl,twocolumn,superscriptaddress,showpacs,10pt]{revtex4}
\usepackage{graphicx}
\usepackage{amssymb}

\begin{document}

\title{The angular momentum of a magnetically trapped atomic condensate}
\author{P. Zhang}
\affiliation{School of Physics, Georgia Institute of Technology, Atlanta, Georgia 30332,
USA}
\author{H. H. Jen}
\affiliation{School of Physics, Georgia Institute of Technology, Atlanta, Georgia 30332,
USA}
\author{C. P. Sun}
\affiliation{Institute of Theoretical Physics, Chinese Academy of Science, Beijing
100080, China}
\author{L. You}
\affiliation{School of Physics, Georgia Institute of Technology, Atlanta, Georgia 30332,
USA}
\affiliation{Center for Advanced Study, Tsinghua University, Beijing 100084, People's
Republic of China}
\date{\today}

\begin{abstract}
For an atomic condensate in an axially symmetric magnetic trap,
the sum of the axial components of the orbital angular momentum
and the hyperfine spin is conserved. Inside an Ioffe-Pritchard
trap (IPT) whose magnetic field (B-field) is not axially
symmetric, the difference of the two becomes surprisingly
conserved. In this paper we investigate the relationship between
the values of the sum or difference angular momentums for an
atomic condensate inside a magnetic trap and the associated gauge
potential induced by the adiabatic approximation. Our result
provides significant new insight into the vorticity of
magnetically trapped atomic quantum gases.
\end{abstract}

\maketitle


A magnetic trap constitutes one of the key enabling technologies for the
recent successes in atomic quantum gases \cite{bergman}. The most commonly
employed magnetic traps includes the quadrupole trap (QT) as in the magneto-optical-trap
(MOT) configuration \cite{mot} and the Ioffe-Pritchard trap (IPT) \cite%
{Pritchard}. The direction of the B-field forming the magnetic trap is
generally a function of the spatial position. For a trapped atom, its
hyperfine spin adiabatically follows the changing B-field direction,
and the atom remains aligned (or anti-aligned) with respect to the local B-field. As a
result of the adiabatic approximation, the center of mass motion of a
magnetically trapped atom experiences an induced gauge potential from the
changing B-field \cite{sun and other,Ho-Shenoy}.

In a variety of magnetic traps, e.g., in a QT, the B-field is
invariant with respect to rotations along a fixed $z$-axis. As the
cse of single particle dynamics \cite{bergman-single}, such a
$\mathrm{SO}(2)$ symmetry leads to the conservation of the
$z$-component $J_{z}$ ($=L_z+F_z$) of the total angular momentum
or the sum of the $z$-components of the atomic spatial angular
momentum $\vec L$ and the hyperfine spin $\vec F$. A different
symmetry exists for an IPT giving rise to a corresponding
conserved quantity $D_{z}$ ($=L_z-F_z$), the difference of $L_z$
and $F_z$. To our knowledge, this surprising property has never
before been identified explicitly. We thus feel obliged to present
this letter because it significantly affects the vortical
properties of a global condensate ground state in a magnetic trap.

Our work is focused on a detailed investigation of the
relationship between the gauge potential and the associated values
of $J_{z}\ (D_{z})$ in a magnetic trap. For a spin-$F$ condensate,
due to the appearance of the adiabatic gauge potential, the
possible values of $J_{z}$ or $D_{z}$ are restricted to a definite
region $[-F,F]$. The gauge potential of our formulation is
directly related to the effective trap rotation studied earlier in
Ref. \cite{Ho-Shenoy}. While Ho and Shenoy
mainly studied the orbital angular momentum component in an IPT \cite%
{Ho-Shenoy}, instead we concentrate on the conserved quantity, the sum ($J_z$) or
difference ($D_z$) for a QT or a IPT.


This paper is organized as follows. We first consider
a spin-$1$ condensate in a QT. Making use of
an effective energy functional appropriate for the adiabatic
approximation \cite{Ho-Shenoy}, we prove $J_{z}\in[-1,1%
]$ with the actual value determined by the
angle between the $z$-axis and the direction of the B-field. We then
generalize to the spin-$F$ case.
Finally, our result is extended to an IPT.

The Hamiltonian of a spin-$1$ atomic condensate with $N$ atoms in
a magnetic trap is $H=H_{S}+H_{I}$ with the single atom part
\begin{eqnarray}
H_{S}=\int\hat{\psi}^{\dagger}(\vec{r})\left[-{\frac{\nabla^{2}}{2M}}+\mu_{%
\mathrm{B}}g_{F}B(\vec{r})\vec{F}\cdot\hat{n}(\vec{r})\right]\hat{\psi}(\vec{%
r})d\vec{r} ,  \nonumber
\end{eqnarray}
and the atom-atom interaction Hamiltonian
\begin{eqnarray}
H_{I}=\sum_{m,n,p,q}\int\hat{\psi}^{(z)\dagger}_{m}(\vec{r})\hat{\psi}%
^{(z)\dagger}_{n}(\vec{r}^{\prime}) V^{mn}_{pq}(\vec{r},\vec{r}^{\prime})%
\hat{\psi}^{(z)}_{p}(\vec{r})\hat{\psi}^{(z)}_{q}(\vec{r}^{\prime})d\vec{r}d%
\vec{r}^{\prime}.  \nonumber
\end{eqnarray}
$\hat{\psi}(\vec{r})=[\hat{\psi}^{(z)}_{-1}(\vec{r}),\hat{\psi}^{(z)}_{0}(%
\vec{r}),\hat{\psi}^{(z)}_{1}(\vec{r})]^{\mathrm{T}}$ denotes the
annihilation field operator for the $z$-quantized $F_{z}$-component of $%
m,n,p,q=0,\pm 1$. $M$ is the atomic mass, and $\mu_{\mathrm{B}}$ is the Bohr
magneton. $B(\vec{r})$ and $\hat{n}(\vec{r})$ denote the strength and
direction of the local B-field. $\hbar=1$ is assumed. The \textit{Lande} $g$
factor is $g_{F=1}=-1/2$.

Within the mean field approximation, the field operator $\hat{\psi}(\vec{r})$
is replaced by its average $\langle\hat{\psi}(\vec{r})\rangle$. To introduce
the adiabatic approximation, we define a group of normalized scalar wave functions $%
\varphi_{u}({\vec r})$:
\begin{eqnarray}
\langle\hat{\psi}(\vec{r})\rangle= \sum_{b=0,\pm 1}\sqrt{N}\xi^{B}(b,\vec{r}){\varphi%
}_{b}(\vec{r}),  \label{meanfield}
\end{eqnarray}
where $\xi^{B}(b,\vec{r})$ is the eigenstate of the $B$-quantized
spin component $\vec{F}\cdot\hat{n}(\vec{r})$ with eigenvalue $b$,
satisfying the
relations $\vec{F}\cdot\hat{n}(\vec{r})\,\xi^{B}(b,\vec{r})=b\,\xi^{B}(b,%
\vec{r})$ and $\xi^{B\dagger}(b,\vec{r})\xi^{B}(b^{\prime},\vec{r}%
)=\delta_{b,b^{\prime}}$. In the $z$-quantized representation, it takes the
form $\xi^{B}(b,\vec{r})=[\xi^{B}_{-1}(b,\vec{r}),\xi^{B}_{0}(b,\vec{r}%
),\xi^{B}_{1}(b,\vec{r})]^{\mathrm{T}}$. In this study, it is
important to distinguish $\xi^{B}(b,\vec{r})$ from the eigenstates
$\xi_{z}(0,\pm 1)$ of $F_{z}$ with eigenvalues $0,\pm 1$. In
explicit form, we have $%
\xi_{z}(-1)=[1,0,0]^{\mathrm{T}}$, $\xi_{z}(0)=[0,1,0]^{\mathrm{T}}$, and $%
\xi_{z}(1)=[0,0,1]^{\mathrm{T}}$.

A magnetic dipole precesses around the direction of a B-field. Majorona
transitions between different $\xi^{B}(b,\vec{r})$ states can be neglected
when $B(\vec{r})$ is large enough. Thus the atomic hyperfine spin
adiabatically freezes in the low-field seeking state $\xi^{B}(-1,\vec{r})$
during the trapped center of mass motion, and $\varphi_{-1}(\vec{%
r})=\varphi(\vec{r})$ and $\varphi_{0,+1}(\vec{r})=0$. Similarly the $z$%
-quantized mean field becomes
\begin{eqnarray}
\hat{\psi}(\vec{r})\approx\langle\hat{\psi}(\vec{r})\rangle=\sqrt{N}\xi^{B}(-1,\vec{r})\varphi(\vec{r}),
\label{wavefunction}
\end{eqnarray}
with $\varphi(\vec{r})$ a B-quantized scalar function.

Substituting Eq. (\ref{wavefunction}) into the expression of $H$,
we can obtain the expression of the condensate energy $E_{\rm ad}$
as a functional of the scalar wave function $\varphi(\vec{r})$:
$E_{\mathrm{ad}}[\varphi]=N\int\varphi^{\ast}(\vec{r})\mathcal{E}_{%
\mathrm{ad}}\varphi(\vec{r})d\vec{r}$. Here
$\mathcal{E}_{\mathrm{ad}}$ defined as \cite{Ho-Shenoy,sun and
other}:
\begin{eqnarray}
\mathcal{E}_{\mathrm{ad}}=\frac{[-i\nabla+\vec{A}(\vec{r})]^{2}}{2M}%
+\mu_{B}B(\vec{r})+W+V_{o} +\frac{g_{2}}{2}|\varphi|^{2}.\   \label{Erm}
\end{eqnarray}
The gauge potential induced by the adiabatic approximation is $\vec{A}(\vec{r%
})=-i\xi^{B\dagger}(-1,\vec{r})\nabla\xi^{B}(-1,\vec{r})$ and $%
W=(|\nabla\xi^{B\dagger}(-1,\vec{r})\nabla\xi^{B}(-1,\vec{r})|-\vec{A}\cdot%
\vec{A})/(2M)$. The trap potential $\mu_{B}B(\vec{r})$ is
augmented by $V_{o}(\rho,z)$ from other sources, e.g., an optical
potential with rotational symmetry along the $z$-axis. Under the
approximation of contact pseudo-potentials, trapped atoms collide
in the same spin aligned state. Thus the $g_{2}$ term is
proportional to the scattering length $a_{2}$ in the total spin
channel $F_{\mathrm{tot}}=2$.

We show below that the gauge potential ${\vec A}(\vec{%
r})$ constrains the values of $J_{z}$ for a condensate ground state.
In cylindrical coordinate $(\rho,\phi,z)$, the B-field
of a QT is expressed as $\vec{B}(\vec{r})=B^{\prime}(\rho\hat{e}_{\rho}-2z%
\hat{e}_{z})$. An optical potential $V_{o}(\rho,z)$ is introduced
to push atoms away from the region of small $B(\vec r)$,
eliminating the deadly Majorona transitions
\cite{mit,opticalplug}. Because this B-field is cylindrically
symmetric, $\vec{F}\cdot\hat{n}(\vec{r})$ commutes with the
$z$-component of the total atomic angular momentum
$J_{z}=L_{z}+F_{z}$. This allows us
to choose $\xi^{B}(-1,\vec{r})$ as the common eigenstate of $\vec{F}\cdot%
\hat{n}(\vec{r})$ and $J_{z}$ with any possible eigenvalues. In this study
we choose $\xi^{B}(-1,\vec{r})$ for convenience to satisfy
\begin{eqnarray}
J_{z}\xi^{B}(-1,\vec{r})=0.  \label{xi}
\end{eqnarray}
This constraint on $J_{z}$ limits the
respective values for $L_{z}$ or $F_{z}$, in a sense
equivalent to a gauge choice. The actual value of $J_z$ for the
ground state of a condensate is determined by appropriate system
parameters.
Explicitly, a simple rotation gives $\xi^{B}(-1,\vec{r})=%
\exp[i\phi] \exp[-i\vec{F}\cdot\hat{e}_{\phi}\beta(\rho,z)]\xi_{z}(-1)$.
The angle between the $z$-axis and $\hat{n}(\vec{r})$
is $\beta(\rho,z)=\arccos[-2z/\sqrt{\rho^{2}+4z^{2}}]$.

The $B$-quantized ground state scalar mean field wave function is denoted as
$\varphi_{g}(\vec{r})$, determined from a minimization of $E_{\mathrm{ad}%
}[\varphi]$. The cylindrical symmetry of both $\vec{B}(\vec{r})$ and $V_{o}$
assures $E_{\mathrm{ad}}[\varphi(\rho,\phi,z)]=E_{\mathrm{ad}%
}[\varphi(\rho,\phi+\theta,z)]$ for any $\theta$. Due to this $\mathrm{SO}(2)
$ symmetry, if there exists only one normalized ground state (up to global
phase factors) as in the situation considered here for a condensate, it has
to be a common eigenstate for all rotation operators $\exp[%
\theta\partial/(\partial\phi)]$, i.e., an eigenstate of $i\partial/(\partial%
\phi)$. Thus, we take $\varphi_{g}(\vec{r})=%
\tilde{\varphi}_{g}(\rho,z)\exp(is\phi)$. On substituting into Eq. (\ref%
{wavefunction}), we obtain the $z$-quantized mean field $\langle\hat{\psi}({%
\vec r})\rangle_{g}$ for the ground state
\begin{eqnarray}
\langle\hat{\psi}(\vec{r})\rangle_{g}=\sqrt{N}\xi^{B}(-1,\vec{r})\tilde{\varphi}%
_{g}(\rho,z)\exp(is\phi),  \label{ground}
\end{eqnarray}
which is an eigenstate of $J_{z}$ with an eigenvalue $s$, i.e.,
\begin{eqnarray}
J_{z}\langle\hat{\psi}(\vec{r})\rangle_{g}=s\langle\hat{\psi}(\vec{r}%
)\rangle_{g},  \label{eigen}
\end{eqnarray}
because of Eq. (\ref{xi}).

Equation (\ref{ground}) and the
expansion $\langle\hat{\psi}(\vec{r})\rangle_{g}=\sum_{m=0,\pm 1}\langle\hat{%
\psi}^{(z)}_{m}({\vec r})\rangle_{g}\xi_{z}(m)$ gives
\begin{eqnarray}
\langle\hat{\psi}^{(z)}_{m}\rangle_{g}=\sqrt{N}\xi^{B}_{m}(-1,{\vec r})\varphi_{g}({%
\vec r})=b_{m}(\rho,z)\tilde{\varphi}_{g}(\rho,z)e^{i(s-m)\phi},
\label{component}
\end{eqnarray}
with $b_{m}=\xi^{\dagger}_{z}(m)e^{-iF_{y}\beta(\rho,z)}\xi_{z}(-1)$.
The individual spin components of state Eq. (\ref{component})
naturally carry topological windings as a
direct result of the conservation of $J_{z}$.

To determine the value of $s$ or $J_{z}$ for the ground
state, we first compute the gauge
potential $\vec{A}(\vec{r})$ for a QT. With the expression of $\xi ^{B}(-1,%
\vec{r})$, we find $\vec{A}(\vec{r}%
)=\cos \beta (\rho ,z)\hat{e}_{\phi }/\rho$. Using the expressions for $%
\vec{A}({\vec{r}})$, $\mathcal{E}_{\mathrm{ad}}$, and $E_{\mathrm{ad}}$, it
is easy to show that for a scalar wave function $\tilde{\varphi}_{g}(\rho
,z)\exp (im\phi )$ with any integer $m$, the energy functional $E_{\mathrm{ad%
}}$ satisfies
\begin{eqnarray}
E_{\mathrm{ad}}[\tilde{\varphi}_{g}(\rho ,z)e^{im\phi }]=E_{\mathrm{ad}}[%
\tilde{\varphi}_{g}]+\Delta E_{m}[\tilde{\varphi}_{g}],\label{E}
\end{eqnarray}
with $E_{m}[\tilde{\varphi}_{g}]=\int d\vec{r}|\tilde{\varphi}_{g}^{\ast
}(\rho ,z)|^{2}[(m^{2}+4m\cos \beta )/(2M\rho ^{2})]$. In addition to the
centrifugal term proportional to $m^{2}$, a term linear in $m$ appears due
to the $\vec{A}\cdot \nabla $ term in $\mathcal{E}_{\mathrm{ad}}$.

In the following we show that the above linear term is
important for the value $s$, which we determine with a variational
approach. Because $E_{\mathrm{ad}}$ takes its minimal value in the state $%
\tilde{\varphi}_{g}\exp [is\phi ]$, we have $E_{\mathrm{ad}}[\tilde{\varphi}%
_{g}e^{is\phi }]\leq E_{\mathrm{ad}}[\tilde{\varphi}_{g}e^{i(s\pm 1)\phi }]$%
. Together with Eq. (\ref{E}), we find the necessary condition satisfied by $%
s$: $\Delta E_{s}[\tilde{\varphi}_{g}]\leq \Delta E_{s\pm 1}[\tilde{\varphi}%
_{g}]$ or $|s+C|\leq 1/2$, where the coefficient $C$ is defined as
\[
C=\frac{\int d\vec{r}|\tilde{\varphi}_{g}^{\ast }(\rho ,z)|^{2}[\cos \beta
(\rho ,z)/\rho ^{2}]}{\int d\vec{r}|\tilde{\varphi}_{g}^{\ast }(\rho
,z)|^{2}[1/\rho ^{2}]}.
\]%
When the correlation between $\cos \beta (\rho ,z)$ and $\rho ^{-2}$ is
neglected, the factor is approximated by $C\approx \int d\vec{r}|%
\tilde{\varphi}_{g}^{\ast }(\rho ,z)|^{2}\cos \beta (\rho ,z)$,
i.e., the expectation value of $\cos \beta (\rho ,z)$ for state
$\tilde{\varphi}_{g}$. Since $|C|<1$, we have $|s+C|\geq |s|-1$,
which is the same as $|s|\leq 1$ or $s\in \lbrack -1,1]$. Without
the gauge
potential $\vec{A}(\vec{r})$, we would always have $\Delta E_{m}[\tilde{%
\varphi}_{g}]=m^{2}\int d\vec{r}|\tilde{\varphi}_{g}^{\ast }(\rho
,z)|^{2}(2M\rho ^{2})^{-1}\geq 0$. Thus the value of $s$ definitely would be
zero. Therefore, the appearance of a non-zero valued $s\in \lbrack -1,1]$
arises due to the induced gauge potential.

We now generalize our result to atoms with an
arbitrary $F$ and inside any axially symmetric B-fields. Analogously we can
prove that the value $s$ of $J_{z}$ in the ground state satisfies the
necessary condition
\begin{eqnarray}
|s-\eta_{F}FC|\leq 1/2,  \label{cc-1}
\end{eqnarray}
and $s\in [-F,F]$ with $\eta_{F}=\mathrm{sign}(g_{F})$.
The result of $s\in [-F,F]$ and the conservation
of $J_z$ is independent of the form of the atomic interaction potential.
Although its strength $g_{2}$ does affect
the wave function shape, thus can
influence the value of $s$ through the factor $C$.

The condition Eq. (\ref{cc-1}) also allows for a rough estimate of $L_{z}$.
A straightforward calculation gives $%
\langle L_{z}\rangle =s-\eta _{F}F\int d\vec{r}|\tilde{\varphi}_{g}^{\ast
}(\rho ,z)|^{2}\cos \beta (\rho ,z)$ for the spinor mean field
$\langle \hat{\psi}(\vec{r})\rangle _{g}$. Neglecting the correlation
between $\rho ^{-2}$ and $\cos \beta (\rho ,z)$ as before, the value of $%
\langle L_{z}\rangle $ becomes approximately $s-\eta _{F}FC$, which
lies always in the region $[-1/2,1/2]$ according to Eq. (\ref{cc-1}).
Therefore,
the value $\langle L_{z}\rangle $, or the weighted average
of the winding numbers, is generally a small number, despite
the winding number $s-m$ itself, for the component $\langle \hat{\psi}_{m}^{(z)}(%
\vec{r})\rangle_{g}$, may take any integer in the region $[-2F,2F]$.
We find
$\langle F_{z}\rangle =\eta _{F}F\int d\vec{r}|%
\tilde{\varphi}_{g}^{\ast }(\rho ,z)|^{2}\cos \beta (\rho ,z)$
from the expression of $\langle L_{z}\rangle$, a
qualitative reflection that atomic hyperfine spin is aligned ($%
g_{F}>0$) or anti-aligned ($g_{F}<0$) with respect to the local B-field.

Our result above allows for the direct creation of vortex states in a
quadrupole trapped atomic condensate. For example, assume a spin-$1$
condensate in a QT plus an ``optical plug" \cite{opticalplug} satisfies $%
V_{o}(\rho,z)=V_{o}(\rho,-z)$, then we find $C=0$ and $s=0$ due to the spatial
reflection symmetry about the $x-y$ plane. The ground state components $\langle\hat{\psi}%
^{(z)}_{\pm 1}(\vec{r})\rangle_{g}$ then automatically carry persistent
currents with winding numbers $\mp 1$ according to Eq. (\ref{component}). In
addition, the low field seeking atoms are trapped near the $x$-$y$ plane at $%
z=0$ because $|B(\vec{r})|$ is an increasing function of $z$. The
populations for the three $z$-quantized states, determined by $\xi^{B}_{0}(-1,%
{\vec r})$ and $\xi^{B}_{\pm 1}(-1,{\vec r})$, are of the same
order of magnitudes. Therefore, when a ground state condensate in
the ``plugged" QT
is created, its $\pm 1$ components $\langle\hat{\psi}^{(z)}_{\pm 1}(\vec{r}%
)\rangle_{g}$ are single quantized vortex states and can be directly
resolved with a Stern-Gerlach B-field as used in Ref. \cite{Ketterle2}.

\begin{figure}[tbp]
\includegraphics[width=3.55in]{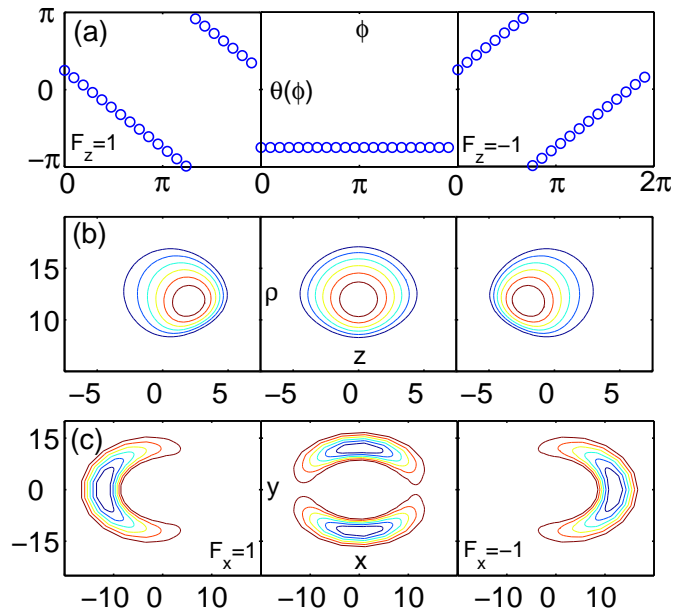}
\caption{(Color online). (a) The phases of the $z$-quantized
components $\langle \hat{\protect\psi}^{(z)}_{1}\rangle$ (left), $\langle
\hat{\protect\psi}^{(z)}_{0}\rangle$ (middle), and $\langle \hat{\protect\psi}%
^{(z)}_{-1}\rangle$ (right panel),
 as functions of the azimuth angle $\protect\phi$;
  (b) the density distributions $|\langle \hat{\protect\psi}%
^{(z)}_{1}\rangle|^{2}$ (left), $|\langle \hat{\protect\psi}%
^{(z)}_{0}\rangle|^{2}$ (middle), and $|\langle \hat{\protect\psi}%
^{(z)}_{-1}\rangle|^{2}$ (right panel) of the $z$-quantized components
as functions of $\protect\rho$ and $z$; (c) the integrated density
distributions $\int|\langle
\hat{\protect\psi}^{(x)}_{1}\rangle|^{2}dz$ (left), $\int|\langle \hat{\protect\psi%
}^{(x)}_{0}\rangle|^{2}dz$ (middle), and $\int|\langle \hat{\protect\psi}%
^{(x)}_{-1}\rangle|^{2}dz$ (right panel) of the $x$-quantized components
as functions of $x$ and $y$. The units for $x$, $y$, $\protect\rho$,
and $z$ in (b) and (c) are all arbitrary.}
\label{fig1}
\end{figure}

The qualitative example above is confirmed by the
numerical solution for a condensate of $5\times 10^{6}$ $^{23}$Na
atoms in a quadrupole plus a plug trap. We take $B^{\prime}=22$
Gauss/cm and $V_{o}=U_{o}\exp[-\rho^{2}/\sigma^{2}]$ with $%
U_{o}=(2\pi)8\times10^{4}$Hz and $\sigma=7.4\mu$m.
The ground state distribution
$p_{i}=\int d\vec{r}|\langle \hat{\psi}^{(z)}_{i}(%
\vec {r})\rangle|^{2}$ is found to be $p_{\pm 1}=27.2\%$ and $p_{0}=45.6\%$.
The phase and density distributions for the three components $\langle%
\hat{\psi}^{(z)}_{0,\pm 1}(\vec{r})\rangle_{g}$ are shown in Fig.
\ref{fig1} (a, b).

We also can expand the ground state $\langle \hat{%
\psi}(\vec{r})\rangle _{g}$ in terms of the eigenstates $\xi _{x}(m)$ of $%
F_{x}$ with eigenvalues $m$: $\langle \hat{\psi}(\vec{r})\rangle
_{g}=\sum_{m=0,\pm 1}\sqrt{N}\langle
\hat{\psi}_{m}^{(x)}(\vec{r})\rangle _{g}\xi
_{x}(m)$. We then immediately note that $\langle \hat{\psi}_{m}^{(x)}(\vec{r}%
)\rangle _{g}$ is a superposition of vortex states with definite
winding numbers $0$ or $\pm 1$, e.g., $\langle
\hat{\psi}_{0}^{(x)}(\vec{r})\rangle
_{g}=(\sqrt{N}/\sqrt{2})[b_{-1}(\rho,z)e^{i\phi
}-b_{1}(\rho,z)e^{-i\phi }]$. The density distribution of $\langle
\hat{\psi}_{0,\pm 1}^{(x)}(\vec{r})\rangle _{g}$ as
shown in Fig. \ref{fig1}(c) clearly illustrates the interference pattern along the $%
\hat{e}_{\phi }$ direction. As is demonstrated in Fig. \ref{fig1}(c),
the middle panel for $|\langle \hat{\psi}_{0}^{(x)}(\vec{r})\rangle|^{2}$
clearly displays the double peak structure along the azimuthal direction,
arising from
the interference of the terms proportional to $e^{i\pm \phi }$.
Thus, if a Stern-Gerlach B-field is used to separate the $x
$-quantized components $\langle \hat{\psi}_{m}^{(x)}(\vec{r})\rangle _{g}$,
a superposition of vortices with different winding numbers would be obtained.

We now extend our result for an axially symmetric magnetic trap to the
widely used IPT whose B-field possesses a different symmetry. In the region
near the $z$-axis, $ \vec{B}(\vec{r})=B^{\prime}[\cos(2\phi)\hat{e}_{\rho}
-\sin(2\phi)\hat{e}_{\phi}+h\hat{e}_{z}]$, the angle $\beta(\rho,z)$ between
the local B-field and the $z$-axis satisfies $\cos\beta(\rho,z)=h/\sqrt{%
\rho^{2}+h^{2}}$. In this case $J_z$ is no longer conserved due to
the lack of the SO(2) symmetry. However, we find that $D_z$
is now conserved because it commutes with $ \vec{F}\cdot\vec{%
B}(\vec{r})$. Therefore, we can select the low field seeking hyperfine spin
state $\xi^{B}(\eta_{F}F,\vec{r})$ as the eigenstate of $D_{z}$ with an
eigenvalue $-\eta_{F}F$, the same spin state as used in \cite{Ho-Shenoy},
again defined through a rotation $\xi^{B}(\eta_{F}F,\vec{r})=\exp[-i\vec{F}%
\cdot\hat{n}_{\bot}\beta]\xi_{z}(\eta_{F}F)$. For $h>0$, we find
the induced
gauge potential becomes $\vec{A}(\vec{r})=-\eta_{F}F(1-\cos\beta(\rho,z))%
\hat{e}_{\varphi}/\rho$.

Adopting the same notation as before we denote
$\langle\hat{\psi}(\vec{r})\rangle_{g}=\sqrt{N}\varphi_{g}(\vec{r}%
)\xi^{B}(\eta_{F}F,\vec{r})$. Interestingly we find $E_{\mathrm{ad}%
}[\varphi(\rho,\phi,z)]=E_{\mathrm{ad}}[\varphi(\rho,\phi+\theta,z)]$
remains satisfied, and the ground state takes
the form $\varphi_{g}=\tilde{\varphi}_{g}(\rho,z)\exp(iu\phi)$. Therefore $%
\langle\hat{\psi}(\vec{r})\rangle_{g}$ is the eigenstate of $D_z$
with an eigenvalue $d=u-\eta_{F}F$, and its components
$\langle\hat{\psi}^{(z)}_{m}(\vec{r})\rangle_{g}=\sqrt{N}b^{\prime}_{m}(\rho,z)e^{i(m+u-%
\eta_{F}F)\phi}$ analogously carry a persistent current with a winding number
$m+u-\eta_{F}F$. Here $b^{\prime}_{m}=\xi^{\dagger}_{z}(m)e^{-iF_{y}\beta}%
\xi_{z}(\eta_{F}F)$. This result is consistent with the ground state
vortex phase diagram for an $F=1$ condensate
found numerically in the $z=0$ plane of an IPT.
The conservation of $D_z$
as found by us, however, calls for a simpler labelling of
each vortex phase because only one of
three integers ($m_1$, $m_0$, $m_{-1}$) is independent,
as with Eq. (15) of  Ref. \cite{sadreev}.

Following the same reasoning as before, we find
\begin{eqnarray}
\big|d+\eta_{F}F(1-C)\big|\leq {1}/{2},  \label{ccc-1}
\end{eqnarray}
for $d\neq \eta_{F}F$, and $d\in [-F,F]$ or the value of
$D_{z}$ in the ground state lies in the region $[-F,F]$.

In an IPT, atoms are trapped near the $z$-axis where the B-field is
essentially along the $z$-axis direction and $\xi^{B}(\eta_{F}F,\vec{r})$ is
approximately the eigenstate $\xi^{z}(\eta_{F}F)$.
$L_{z}$ then is
essentially always zero corresponding to a ground state without a vortex.
The angular momentum difference $D_z$ then becomes $%
d=-\eta_{F}F$.

Several previous proposals \cite{Machida} and experiments \cite{Ketterle} on
creating vortex states unknowingly have used the fact that $\langle\hat{\psi}(\vec{r})%
\rangle_{g}$ in an IPT is an eigenstate of $D_{z}$. For example,
in the experiment of Ref. \cite{Ketterle}, spin-$1$ atoms
initially were prepared in the internal state $\xi^{z}(-1)$
($F_{z}=-1$) with no vorticity in its spatial mode. This
corresponds to $D_{z}=1$. To create a vortex state, the bias field
along the $z$-axis was adiabatically inverted from the $+z$ to the
$-z$
direction. In this process the atomic internal state was changed from $%
\xi^{z}(-1)$ to $\xi^{z}(1)$. If the whole operation is adiabatic, the
commutator $[\vec{F}\cdot\vec{B}(\vec{r},t),D_{z}]=0$ will be maintained.
Consequently $D_{z}$ is conserved. In the end when the
internal state was changed to $\xi^{z}(1)$ ($F_{z}=1$),
its orbital angular momentum became $L_{z}=D_{z}+F_{z}=2$ or a
double vortex spatial mode as was observed \cite{Ketterle}.

In Ref. \cite{Ketterle2} when the bias field is adiabatically switched off,
the direction of the B-field adiabatically changes to lie in the $x$-$y$
plane. $D_{z}$ is again conserved during this process ($=1$). When complete,
the atomic hyperfine spin state is changed from $\xi^{z}(-1)$ to a
superposition of all three components $\xi^{z}(0,\pm 1)$. As a result,
different $F_{z}$ components are then associated
with vortex states with corresponding $L_{z}=D_{z}+F_{z}$.
When the three internal states are separated, two of them are observed to
contain vortices with non-zero winding numbers.

Creating a ground state condensate with $D_{z}\neq -\eta _{F}F$ is
quite challenging inside an IPT. This was considered
quantitatively by Ho and Shenoy \cite{Ho-Shenoy}, who obtained an
approximate gauge potential resembling an effective trap rotation
when expanded to the first order of $\rho$. However, their
expansion easily fails away from the $z$-axis. A numerical
discussion on this challenge is provided in \cite{sadreev}. The
more general constraint of $D_{z}\neq -\eta _{F}F$, i. e., the
necessary condition Eq. (\ref{ccc-1}) of the angular momentum
difference $D_{z}$, was not obtained in \cite{Ho-Shenoy,sadreev}.
From a direct calculation, it can be proved that
 if $C$ is approximated by $\int d\vec{r}|\tilde{\varphi}%
_{g}^{\ast }(\rho ,z)|^{2}\cos \beta (\rho ,z)$ and $d\neq \eta
_{F}F$, we have $\langle L_{z}\rangle =d+\eta _{F}F\int
d\vec{r}|\tilde{\varphi}_{g}^{\ast }(\rho ,z)|^{2}[1-\cos \beta
(\rho ,z)]$, which can never be greater than $1/2$ according to
Eq. (\ref{ccc-1}).

In summary, we have investigated the angular momentum of a magnetically
trapped condensate. Inside an axially
symmetric trap such as a QT, the total angular momentum $J_z$ along the
symmetric $z$-axis is found to be conserved,  while the angular momentum
difference $D_z$ is conserved in an IPT. Both conservation laws reflect the
underlying symmetries of the traps' magnetic fields, and the values of $J_z$
or $D_z$ in the ground states are determined by the gauge potential $%
\vec{A}(\vec{r})$. In the global ground state, the corresponding eigenvalues
of $J_z$ and $D_{z}$ are limited to $%
\in [-F,F]$ with the precise values directly related to the angle between the
local B-field and the $z$-axis.

Our results provide significant insights into the study of
magnetically trapped condensates. The conservation laws we discuss
reveal an important observational consequence: the components $\langle \hat{%
\psi}_{m}^{(z)}(\vec{r})\rangle _{g}$ of a condensate ground state
$\langle \hat{\psi}(\vec{r})\rangle _{g}$ automatically carry
persistent currents with different winding numbers.
Furthermore, according to the conditions Eqs. (\ref{cc-1}) and (\ref{ccc-1}%
), the values of $J_{z}$ or $D_{z}$, or the topological winding
numbers of the spatial wave functions, can be controlled through
the angle $\beta (\rho ,z)$ between the local B-field and the
$z$-axis. We have shown explicitly for a condensate in a QT with
an optical plug, where the atomic populations for the $2F+1$
components have approximately the same order of magnitude.
Therefore, vortex states can be present already in a ground state
condensate without requiring adiabatic operations as in
\cite{Ketterle,Ketterle2}.

We thank Dr. D. L. Zhou, Mr. B. Sun, Profs. C. Raman and W. Ketterle for
helpful discussions. This work is supported by NSF, NASA, and NSFC.

\end{document}